\documentclass[final,1p,times,twocolumn]{elsarticle}

\usepackage{graphics}
\usepackage{graphicx}
\usepackage{epsfig}
\usepackage{color}
\usepackage{amssymb}
\usepackage{amsmath}

\usepackage{color}

\begin{document}

\begin{frontmatter}



\title{Approximation of slow and fast dynamics in multiscale dynamical systems by the linearized Relaxation Redistribution Method}


\author{Eliodoro Chiavazzo \fnref{LAV}\corref{cor1}}
\ead{eliodoro.chiavazzo@polito.it}

\cortext[cor1]{Corresponding author: Department of Energetics, Politecnico di Torino, Corso Duca degli Abruzzi 24, 10129 Turin, Italy. Fax: +39 011 564 4499}

\address[LAV]{Department of Energetics, Politecnico di Torino, Corso Duca degli Abruzzi 24, 10129 Turin, Italy}

\begin{abstract}
In this paper, we introduce a fictitious dynamics for describing the only fast relaxation of a stiff ordinary differential equation (ODE) system towards a stable low-dimensional invariant manifold in the phase-space ({\em slow invariant manifold} - SIM). As a result, the demanding problem of constructing SIM of any dimensions is recast into the remarkably simpler task of solving a properly devised ODE system by stiff numerical schemes available in the literature. In the same spirit, a set of equations is elaborated for local construction of the fast subspace, and possible initialization procedures for the above equations are discussed. The implementation to a detailed mechanism for combustion of hydrogen and air has been carried out, while a model with the exact Chapman-Enskog solution of the invariance equation is utilized as a benchmark.
\end{abstract}


\end{frontmatter}

\section{Introduction}\label{intro}
Models for describing detailed reaction mechanisms of hydrocarbon fuels and biochemical processes in living cells are typical examples of large multiscale dynamical systems \cite{Jamshidi2008,Endy01,Pana2008,GriMech30_long}. In this respect, modern research has to cope with an increasing difficulty mainly in two aspects: First, the number of degrees of freedom is tremendously large making it difficult to obtain a physical understanding of the above phenomena. In addition, computations are often dramatically time consuming due to a wide range of time-scales to be resolved. As a result, methodologies able to tackle the above issues become highly desirable.
The issue of physical understanding is drawing an increasing attention in the realm of kinetic modeling of biological systems with many degrees of freedom \cite{Leb2005PRE,Leb_Skanda2008,Surov2009}.
%
Modern simplification techniques (often referred to as {\em model reduction methods} \cite{ModRedCollection}) are based on a systematic decoupling of the fast processes from the rest of the dynamics, and are typically implemented by seeking a \textit{low dimensional manifold} in the phase-space. Towards this end, several methods have been suggested in the literature \cite{book,CKGB09,MP92,LG91,VALORANI-JCP,Lebiedz2011,KAN2010,Adrover2007JCP} which are based on the following picture (for dissipative systems with unique steady state to be addressed below): Multiscale systems are characterized by a short transient towards a stable low-dimensional manifold in the phase space, known as the slow invariant manifold (SIM). The subsequent dynamics is slower and proceeds along the manifold, until a steady state is reached.

Recently, the Relaxation Redistribution Method (RRM) has been proposed \cite{PHD_THESIS_EC} and implemented in realistic combustion mechanisms for hydrogen \cite{CKPRE2010} and methane mixtures \cite{CAV2011}. RRM has been introduced as a particularly efficient scheme to implement the film equation of dynamics (see section \ref{background} below), which can be used to construct the SIM and adaptively choose the minimal description of complex multiscale systems \cite{CKPRE2010,ChKa2011}.
In the latter References, the minimal description is understood as the minimal dimension of a convergent SIM (by RRM). For completeness, we stress that an alternative approach for the adaptive simplification of multiscale systems is the G-Scheme in \cite{Gscheme}.

In the present work, {following} the basic idea behind the RRM, we derive a set of ordinary differential equations which approximate the RRM dynamics (here, referred to as {\em governing equations of the linearized RRM}). A remarkably easy implementation of the latter method for constructing SIM in any dimensions is then proposed.

This manuscript is organized as follows. In section \ref{background}, the problem of model reduction is introduced and the notions of both invariance equation and film equation are briefly reviewed. The governing equations of the linearized RRM are presented in section \ref{slow.equations}, where the link between the {presented} method and other approaches (i.e. direct solution of invariance and film equations \cite{book}, ILDM \cite{MP92}, CSP \cite{LG91}) is shortly discussed. A novel algorithm for approximating the fast subspace is introduced in section \ref{fast.algorithm}. {The accuracy in describing the SIM by governing equations of the linearized RRM at steady state is discussed in section \ref{steady.state.accuracy}, whereas}
a possible initialization of {them is proposed} in section \ref{inizio}. The suggested {linearized RRM} is tested in section \ref{BEN-EX}, while conclusions are drawn in section \ref{conclusions}.
\section{Background}\label{background}
Let an autonomous system of ordinary differential equations
\begin{equation}\label{ODEgen}
\frac{{dY}}{{dt}} = \left[ {\begin{array}{*{20}c}
   {f_1 \left( Y \right)}  \\
    \vdots   \\
   {f_n \left( Y \right)}  \\
\end{array}} \right] = f\left( Y \right),
\end{equation}
describe the time evolution of a state $Y=[c_1,...,c_n]^T$ in the phase-space $U$, where $n$ is the dimension of $U$ and the superscript $^T$ denotes transposition. {\em Model reduction} techniques enable the construction of a simplified ODE system 
\begin{equation}\label{ODEred}
\frac{d \xi}{dt}=f' \left( \xi \right),
\end{equation}
where the state $ \xi = \left[ \xi_1,...,\xi_q \right]^T $ belongs to a {\em reduced} space $\Xi$ with dimension $q<<n$, and evolves in time according to the slow dynamics of system (\ref{ODEgen}). 
%
\subsection{Slow invariant manifolds}
By analogy with classical thermodynamics, a reduced model (\ref{ODEred}) represents a {\em macroscopic description} of a physical phenomenon (given by (\ref{ODEgen})) where various processes with disparate timescales occur.

Formally, the link between the microscopic world and the corresponding macroscopic description can be established by resorting to the notion of slow invariant manifold (SIM). In other words, the reduced (macroscopic) dynamics (\ref{ODEred}) occurs along a $q$-dimensional SIM, $\Omega^{SIM}$, embedded in the phase-space $U$. 
Thus, through $\Omega^{SIM}$, it is possible to pick up the most likely microstate among all the possible ones which are consistent with a macrostate characterized by the {\em macroscopic observables} $\xi \in \Xi$ (see also \cite{LebiedzEntropy}).
In the following, by model reduction, we mainly refer to constructive methods of both the slow and fast subspaces, and we assume that an arbitrary manifold $\Omega$ can be defined (at least locally) by means of a mapping
\begin{equation}\label{mapping}
\mathcal{F}: \Xi \rightarrow U.
\end{equation}
By definition, $\Omega^{SIM} \subset U$ is {\it invariant} with respect to the system (\ref{ODEgen}) if inclusion $Y(t_0) \in \Omega $ implies that
$Y(t) \in \Omega $ for all future times $t > t_0$.
Equivalently, if the tangent space $T_{\emph{\textbf{y}}}$ to $\Omega$ is defined at $Y$, invariance requires: $f(Y) \in T_{\emph{\textbf{y}}}.$

In order to transform the latter condition into an equation, it proves convenient to introduce projector operators. For any subspace $T_{\emph{\textbf{y}}}$, let a
projector $P$ onto $T_{\emph{\textbf{y}}}$ be defined with image ${\rm im} P
=T_{\emph{\textbf{y}}}$ and $P^2 = P$. Then the invariance condition
can be expressed as:
\begin{equation}\label{invEq}
\Delta=(I - P)f=0,
\end{equation}
where $\Delta$ is often called {\it defect of invariance} \cite{book}, and $I$ represents the identity matrix.

It is worth stressing that, although the notion of invariance discussed above is relatively straightforward, {\em slowness} instead is much more delicate. We just notice that invariant manifolds are not necessarily suitable for model reduction (e.g., all semi-trajectories are, by definition, one-dimensional (1D) invariant manifolds). 
For singularly perturbed systems, the notion of slow invariant manifold has been defined in the framework of the geometric singular perturbation theory by Fenichel \cite{Fenichel1979}.
However, we should also point out that in general the different methodologies proposed in the literature for model reduction purposes are based on different objects. {For instance, it is known that the rate controlled constrained equilibrium (RCCE) manifold \cite{KeG71,HBMK98} typically does not even fulfill the invariance condition (\ref{invEq}), whereas other methods (see, e.g., \cite{book,PHD_THESIS_EC}, \cite{RenPope06}, \cite{MP92}) attempt the construction of invariant objects (with different accuracy).}

Here, we follow the rationale behind the Method of Invariant Manifold (MIM) \cite{book,ChGoKa07}, where slowness is understood as {\em stability} (see also chapter 4 of \cite{book}), so that a SIM is the {\it stable stationary} solution of a relaxation process ({\em film equation})
\begin{equation}\label{film} 
\frac{{\rm d} \mathcal{F}(\xi)}{{\rm d} t}=(I - P) f.
\end{equation}
%
%
We notice that the projector operator $P$ introduces first order spatial derivatives (with respect to the manifold parameters $\xi$). Therefore, (\ref{invEq}) and (\ref{film}) are partial differential equations (see also \cite{book,Nafe02}) whose unknown is the function $\mathcal{F}$, which is conveniently utilized for a parametric representation of the manifold $\Omega$, with $\Omega$ being the image of $\mathcal{F}$: $\Omega = \mathcal{F}(\xi), \; \xi \in \Xi$.

Several numerical schemes have been suggested in the literature for solving Eqs. (\ref{invEq}) and (\ref{film}): The Newton method with incomplete linearization and the relaxation method in \cite{book}, the semi-implicit scheme in \cite{Nafe02,BM07} represent a few examples. The latter approaches aiming at the direct solution of both the invariance condition and film equation are often hindered by severe numerical (Courant type) instabilities \cite{Cou67,GKZ04,ChGoKa07}. Toward the end of overcoming the latter issues, the Relaxation Redistribution Method (RRM) has been recently introduced \cite{ChKa2011,CKPRE2010} (see also Fig. \ref{schizzo}). In the following, exploiting the rationale behind the RRM, we devise a set of ordinary differential equations approximating the dynamics of the film equation (\ref{film}) in a neighborhood of a fixed macrostate $\bar \xi \in \Xi$.  

\section{Linearized Relaxation Redistribution Method (RRM)}\label{main.alg}
\subsection{Slow subspace}\label{slow.equations}
\begin{figure}
	\centering
		\includegraphics[trim = 60mm 9mm 30mm 35mm, clip, width=0.7\textwidth]{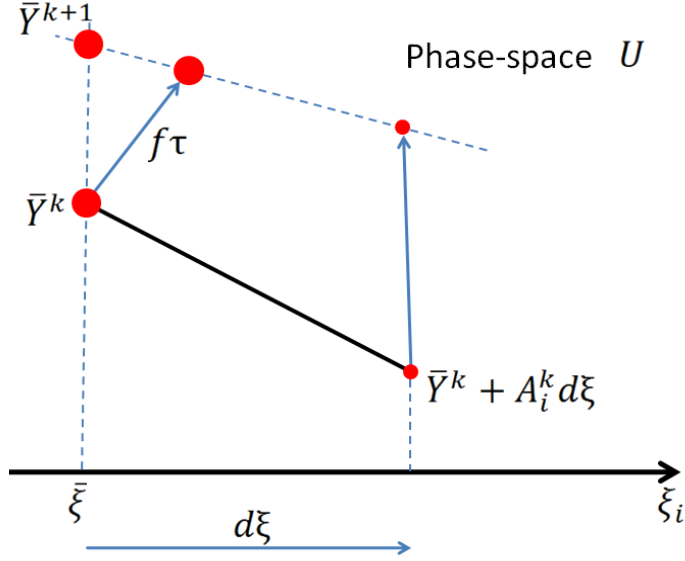}
	\caption{(Color online). Schematic representation of the basic idea behind the Relaxation Redistribution Method (RRM). In a small neighborhood of the pivot $\bar Y^k$ (large circles), a linear approximation of the SIM at the iteration $k$ is considered. With the aim of driving the pivot towards the SIM, RRM \cite{ChKa2011,CKPRE2010,CAV2011} prescribes an updating rule $\bar Y^k \rightarrow \bar Y^{k+1}$ as schematically sketched above (small circles represent neighbors of the pivot). Here, an ODEs system (\ref{Y.evol}) whose dynamics approximates the latter updating rule is suggested and tested.}\label{schizzo}
\end{figure}
Let the dynamical system (\ref{ODEgen}) be characterized by a hierarchy of time scales, and let $\tau$ be of the order of the fastest scale of (\ref{ODEgen}). Let the $q \times n$ matrix $B$ and its $j$-th row $B_j$ 
\begin{equation}\label{Bmatrix}
B = \left[ {\begin{array}{*{20}c}   {b_{11} } &  \cdots  & {b_{1n} }  \\    \vdots  &  \ddots  &  \vdots   \\   {b_{q1} } &  \cdots  & {b_{qn} }  \\\end{array}} \right],\;B_j  = \left[ {b_{j1} ,...,b_{jn} } \right],
\end{equation}
define a linear mapping from the phase-space $U$ into a reduced space $\Xi$ of dimension $q << n$: 
\begin{equation}\label{micro2macro}
BY = \xi ,\quad \xi  = \left[ {\begin{array}{*{20}c}
   {\xi _1 }  \\
    \vdots   \\
   {\xi _q }  \\
\end{array}} \right] \in \Xi ,
\end{equation}
such that a macrostate $\xi$ can be associated with any microstate $Y$ via the (\ref{micro2macro}).
In the following, we develop an {\em iterative} methodology for refining an initial approximation of the SIM in a vicinity of a given macrostate $\bar \xi \in \Xi$. To this end, at each iteration $k$, we assume that, in a neighborhood of $\bar \xi$, the SIM is approximated by an affine linear mapping, $\mathcal{F}^k: \Xi \to U$, of the form:
\begin{equation}\label{linearized.manifold}
\mathcal{F}^k : \quad Y  = A^k   \xi  + l^k , \quad \mathcal{F}^k \left(\bar \xi \right) = \bar Y^k,
\end{equation}
where $A^k$ and $l^k$ are a $n \times q$ matrix and a $n \times 1$ column vector, respectively, such that: 
\begin{equation}\label{AAl}
A^k  = \left[ {\begin{array}{*{20}c}   {a_{11}^k } &  \cdots  & {a_{1q} ^k}  \\    \vdots  &  \ddots  &  \vdots   \\   {a_{n1}^k } &  \cdots  & {a_{nq}^k }  \\\end{array}} \right],\;A_i^k  = \left[ {\begin{array}{*{20}c}   {a_{1i} ^k}  \\    \vdots   \\   {a_{ni} ^k}  \\\end{array}} \right],\;\;l^k  = \left[ {\begin{array}{*{20}c}   {l_1 ^k}  \\    \vdots   \\   {l_n ^k}  \\\end{array}} \right].
\end{equation}
Notice that, the over-bar denotes the pivot ($\bar Y^k$ in Fig. \ref{schizzo}) at an arbitrary iteration $k$ along with the corresponding macrostate $\bar \xi$.

For the state $\bar Y^k=\mathcal{F}^k(\bar \xi)$ belongs to the space defined by the linear function in (\ref{linearized.manifold}), the column vector $l^k$ satisfies:
\begin{equation}\nonumber
l^k  = \bar Y^k  - A^k \bar \xi ,\quad \bar \xi  = B\bar Y^k.
\end{equation}
Assuming the existence of a $q$-dimensional SIM, $\Omega^{SIM}$, we aim at devising a procedure for updating 
the linear mapping (\ref{linearized.manifold}):
\begin{equation}\label{linearized.manifold.up}
\mathcal{F}^{k+1} : \quad Y  = A^{k+1}   \xi  + l^{k+1} ,
\end{equation}
such that (\ref{linearized.manifold.up}) describes $\Omega^{SIM}$ with a better accuracy than (\ref{linearized.manifold}) in a neighborhood of $\bar \xi$.
Toward this end, we follow the rationale behind the Relaxation Redistribution Method (RRM) introduced in \cite{PHD_THESIS_EC,CKPRE2010}. We stress that, at any iteration $k$, the manifold is described by the mapping $\mathcal{F}^{k}$ with the $(q+1)$ microstates $\bar Y^k$, $\bar Y^k + A^k_i d \xi$ belonging to the affine subspace (\ref{linearized.manifold}).

In Fig. \ref{schizzo}, we pictorially show the relaxation of  $\bar Y^k$ (large circle) and one of its $q$ neighbors (small circle) 
\begin{equation}\nonumber
\bar Y^k + A^k_i d\xi, \quad i=1,...,q,
\end{equation}
(all in the space defined by $\mathcal{F}^k$) 
toward $\Omega^{SIM}$ during the time $\tau$, where $d \xi$ is a small parameter. {Notice that, owing to arbitrariness in picking the $i$-th neighbor, for simplicity we make the choice $d\xi_i=d\xi, \; \forall i$.}
According to the RRM algorithm \cite{CKPRE2010}, $\mathcal{F}^{k+1}$ describes the subspace defined by the set of $q+1$ relaxed states. 
The updated points can be written as:
\begin{equation}\label{relaxed.points}
\begin{array}{l}
\bar Y^k + f \left(\bar Y^k \right) \tau, \\
\bar Y^k + A^k_i d \xi +  f \left(\bar Y^k + A^k_i d \xi \right)  \tau, \quad i=1,...,q,
\end{array}
\end{equation}
which represent the advance in time of the $(q+1)$ points $\bar Y^k$, $\bar Y^k + A^k_i d \xi$ during a period $\tau$, according to an explicit Euler scheme.
Upon linearization of the right-hand side of (\ref{ODEgen}), $f\left(Y + dY \right) \approx f\left(Y \right)+J\left(Y \right) dY$,
%
%
(\ref{relaxed.points}) take the approximate form:
\begin{equation}\label{relaxed.points.approx}
\begin{array}{l}
\bar Y^k + f \left(\bar Y^k \right) \tau, \\
\bar Y^k + A^k_i d \xi +  f \left(\bar Y^k\right) \tau + J\left( \bar Y^k \right) A^k_i d \xi   \tau, \quad i=1,...,q
\end{array}
\end{equation} 
with $J=J\left( \bar Y^k \right)=\left[ \partial f_i/ \partial Y_j \right]$ denoting the Jacobian matrix evaluated at $\bar Y^k$.
Thus, a set of $q$ vectors spanning the linear space described by $\mathcal{F}^{k+1}$ reads as follows:
\begin{equation}\label{vectors.updated}
 A_i^k + J \left(\bar Y^k\right) A_i^k \tau, \quad i=1,...,q.
\end{equation}
The RRM algorithm \cite{CKPRE2010} introduces a {\em fictitious dynamics} such that an arbitrary state $\bar Y^{k}=\mathcal{F}^k \left( \bar \xi \right)$ "moves" to a new location:
\begin{equation}\label{F.updated}
 \bar Y ^{k+1} =\mathcal{F}^{k+1}\left( \bar \xi \right)= \sum_{i=1}^{q} \bar \alpha_i \left[ A_i^k + J \left( \bar Y^k \right) A_i^k \tau \right]  + \bar Y^k + f \left( \bar Y^k \right) \tau,
\end{equation}
which obeys the relationship:
\begin{equation}\label{RRM.condition}
\bar \xi_j=B_j \bar Y^k = B_j \bar Y^{k+1}, \quad j=1,...,q.
\end{equation}
Equations (\ref{RRM.condition}) stipulate that the movement $\bar Y^k \rightarrow \bar Y^{k+1}$ is orthogonal to the parameter space $\Xi$ (see also Fig. \ref{schizzo}) and, upon substitution of (\ref{F.updated}) in the (\ref{RRM.condition}), enable the explicit computation of the variables $\bar \alpha_i$ in (\ref{F.updated}) by solving the linear system:
%
\begin{equation}\label{alpha.coeff}
\left[ {\begin{array}{*{20}c}   {1 + B_1 J A_1^k \tau} &  \cdots  & {B_1 J A_q^k \tau}  \\    \vdots  &  \ddots  &  \vdots   \\   { B_q J A_1^k \tau } &  \cdots  & {1 + B_q J A_q^k \tau}  \\\end{array}} \right] \cdot \left[ {\begin{array}{*{20}c}   {\bar \alpha_1 }  \\    \vdots   \\   {\bar \alpha_q }  \\\end{array}} \right] = - \tau \left[ {\begin{array}{*{20}c}   {B_1 f }  \\    \vdots   \\   {B_q f }  \\\end{array}} \right]
\end{equation}
where due to (\ref{micro2macro}) and (\ref{linearized.manifold}) $B_i A_j^k = \delta_{ij} \; \forall k$, with $\delta_{ij}$ being the Kronecker delta.
In the following, in order to save notation, we assume that both the vector field $f$ and the Jacobian matrix $J$ are computed at $\bar Y^k$; equation (\ref{F.updated}) takes the more compact form:
%
\begin{equation}\label{updating}
\bar Y^{k+1}= \bar Y^k + \left( f - M^k \Phi^{-1}_k B f \right) \tau,
\end{equation}
%
with $\Phi_k$ denoting the $q \times q$ matrix on the left-hand side of (\ref{alpha.coeff}), whereas $M^k$ is a $n \times q$ column matrix defined as follows:
%
\begin{equation}\nonumber
M^k = \left[ {A_1^k + J A_1^k \tau}  ,    ...   ,   {A_q^k + J A_q^k \tau}   \right] = \left( I + J \tau \right) A^k,
\end{equation}
with I denoting the $n \times n$ unit matrix.
%
It is worth stressing that the solution of the equations (\ref{updating}) requires an updating rule for the matrix $A$ in (\ref{AAl}) as well: $A^k \rightarrow A^{k+1}$.
For this purpose, we notice that an arbitrary point $Y$ in the linear space described by the function $\mathcal{F}^{k+1}$ takes the form
%
%
\begin{equation}\label{yetanotherF}
\mathcal{F}^{k+1}: \quad Y = \sum_{i=1}^q \alpha_i \left( A_i^k + J A_i^k \tau \right) + \bar Y^{k+1},
\end{equation}
hence the $j$-th parameter $\xi_j$ corresponding to $Y$ is
%
\begin{equation}\nonumber
\xi_j = \sum_{i=1}^{q} \alpha_i \left( \delta_{ij} + B_j J A_i^k \tau \right) + \bar \xi_j,
\end{equation}
%
or, equivalently, in matrix notation:
\begin{equation}\label{Yxi.mat}
Y = M^k \Lambda + \bar Y^{k+1}, \quad \xi = \Phi_k \Lambda + \bar \xi,
\end{equation}
with $\Lambda = \left[ \alpha_1, ... , \alpha_q \right]^T$. 
Equation (\ref{yetanotherF}) stems from (\ref{F.updated}) where the origin, $\bar Y^k + f (\bar Y^k) \tau$, of the affine subspace has been replaced with $\bar Y^{k+1}$.

Equations (\ref{Yxi.mat}) yield the function $\mathcal{F}^{k+1}$ in the form (\ref{linearized.manifold.up}):
\begin{equation}\nonumber
Y = M^k \Phi ^ {-1}_k \left( \xi - \bar \xi \right) + \bar Y^{k+1},
\end{equation}
where 
\begin{equation}\label{Akp1}
A^{k+1}= M^k \Phi ^ {-1}_k, \quad l^{k+1} = \bar Y^{k+1} - M^k \Phi^{-1}_k \bar \xi. 
\end{equation}
The updating rules (\ref{updating}) and (\ref{Akp1}) can be interpreted as the explicit Euler numerical scheme for solving the following dynamical system: 
\begin{equation}\label{Y.evol}
\frac{{dY}}{{dt}} = f - M  \Phi ^{ - 1} B f, \quad \frac{d A}{dt} = \left[ M \Phi^{-1} - A \right] \tau^{-1}.
\end{equation}
The second equation in (\ref{Y.evol}) can be derived, by analogy with (\ref{updating}), after recasting the first equation in (\ref{Akp1}) as follows:
\begin{equation}\nonumber
A^{k+1}=A^k + \left[ \left(M^k \Phi ^ {-1}_k - A^k \right) \tau^{-1} \right] \tau.
\end{equation}
The above equations (\ref{Y.evol}) are the governing equations of the linearized RRM, which dictate a fictitious temporal evolution of a state $Y$ and a $n \times q$ matrix $A$
\begin{equation}\nonumber
A=\left[A_1,...,A_q\right], \quad A_i = \left[ {\begin{array}{*{20}c}   {a_{1i} }  \\    \vdots   \\   {a_{ni} }  \\\end{array}} \right],
\end{equation}
(defining an affine linear mapping of the form $\mathcal{F}:$ $Y=A \xi + l$) towards the corresponding $q$-dimensional slow invariant manifold $\Omega^{SIM}$ in a neighborhood of a given macroscopic state $\bar \xi$, with
%
\begin{equation}\nonumber
M = \left[ {A_1 + J A_1 \tau}  ,    ...   ,   {A_q + J A_q \tau}   \right], \; \Phi(i,j)= \delta_{ij} + B_i J A_j \tau .
\end{equation}
%
%
{For the sake of clarity, we point out that the dynamics (\ref{Y.evol}) (as well as the RRM dynamics in \cite{CKPRE2010}) is referred to as {\em fictitious} because, unlike the original detailed system (\ref{ODEgen}), no physical or chemical processes are typically described by it.}
%

{Moreover,} we stress that the presence of the time-scale $\tau$ in the right-hand side of the equations in (\ref{Y.evol}) introduces a remarkable stiffness, thus the ODEs (\ref{Y.evol}) typically require state of the art stiff integrators (see, e.g., \cite{ode15s}).
According to the RRM method, the SIM is obtained when the relaxation and redistribution steps balance each other (details can be found in \cite{CKPRE2010}). In the suggested algorithm, the analogous condition is satisfied at the steady state (here denoted as $Y=Y^{ss}$ and $A=A^{ss}$) of the dynamical system (\ref{Y.evol}). Hence, the SIM is given (in a vicinity of $\bar \xi$) by:
\begin{equation}\label{SIM.linear}
 \Omega^{SIM}: \quad Y=A^{ss} \xi + l^{ss}, \quad l^{ss} =  Y^{ss} - A^{ss} \bar \xi.
\end{equation}
%
%
%
%
%
%
%
{We stress that computation of the quantity $l^{ss}$ does not require additional refinements, and it is performed by (\ref{SIM.linear}) (upon convergence of (\ref{Y.evol})) if a linear approximation is to be provided for approximating the mapping (\ref{mapping}) in a neighborhood of $Y^{ss}$.}

It is worth noticing that, inspection of the right-hand side of the first equation in (\ref{Y.evol}) reveals a clear connection between the RRM method introduced in \cite{ChKa2011,CKPRE2010} and the film equation (\ref{film}). In fact, although (\ref{Y.evol}) represents a system of ordinary differential equations whereas (\ref{film}) is a partial differential equation, the former only describes the (\ref{film}) locally in a vicinity of a macrostate $\bar \xi$. In this respect, the projector onto the tangent space of a manifold $\Omega$ takes the explicit form: $P = M  \Phi ^{ - 1} B$. In this respect, the latter operator satisfies the condition of projectors: 
\begin{equation}\nonumber
P^2 = M  \Phi ^{ - 1} B M  \Phi ^{ - 1} B = M  \Phi ^{ - 1} B = P,
\end{equation}
due to the relation $B M = \Phi$.
Similarly to (\ref{film}), the governing equations of the linearized RRM (\ref{Y.evol}) prescribe a composition of two motions: the first one along the detailed dynamics $f$, while the second one along the tangent space of $\Omega$, $ - P f = - M  \Phi ^{ - 1} B f$. Finally, at steady state of (\ref{Y.evol}), the invariance condition (\ref{invEq}) is satisfied:
\begin{equation}\nonumber
f - M  \Phi ^{ - 1} B f = \left(I - M  \Phi ^{ - 1} B \right) f = \left(I - P \right) f = 0.
\end{equation}
The above equation imposes that, on the SIM, the component of dynamics $f$ in the fast subspace vanishes. Since that condition lies at the heart of other popular methods (such as ILDM and CSP \cite{MP92,LG91}), this explains the formal resemblance of (\ref{Y.evol}) (at steady state) to the equations adopted in ILDM and CSP.

%
\subsection{Fast subspace}\label{fast.algorithm}
The methodology proposed in the previous section can be utilized for extracting the slow invariant manifold (i.e. the subspace of slow motions or {\em slow subspace}, for short) with respect to the ODE system (\ref{ODEgen}). Nevertheless, this is only one aspect of model reduction: Computing the {\em fast subspace} is indeed required in order to achieve the complete decomposition of the full dynamics $f$ (slow-fast decomposition). 

We notice that, towards this end, several approaches have been proposed in the literature. For instance, the notion of {\em thermodynamic projector} \cite{book,CKG10,ChK2007} for dissipative systems supported by a thermodynamic Lyapunov function, the spectral decomposition of the Jacobian matrix $J$ \cite{MP92}, and the CSP algorithm \cite{GV06} are some of the most popular examples.

Those methods might be adopted in combination with the above technique (\ref{Y.evol}) as well, for an {\em a posteriori} reconstruction of the fast subspace. However, here in the same spirit of the method presented in section \ref{slow.equations}, we propose an alternative procedure for computing the fast subspace, in a neighborhood of a given macrostate $\bar \xi$, once the linear function (\ref{SIM.linear}) has been computed.
\begin{figure}
	\centering
		\includegraphics[trim = 20mm 9mm 12mm -10mm, clip, width=0.8\textwidth]{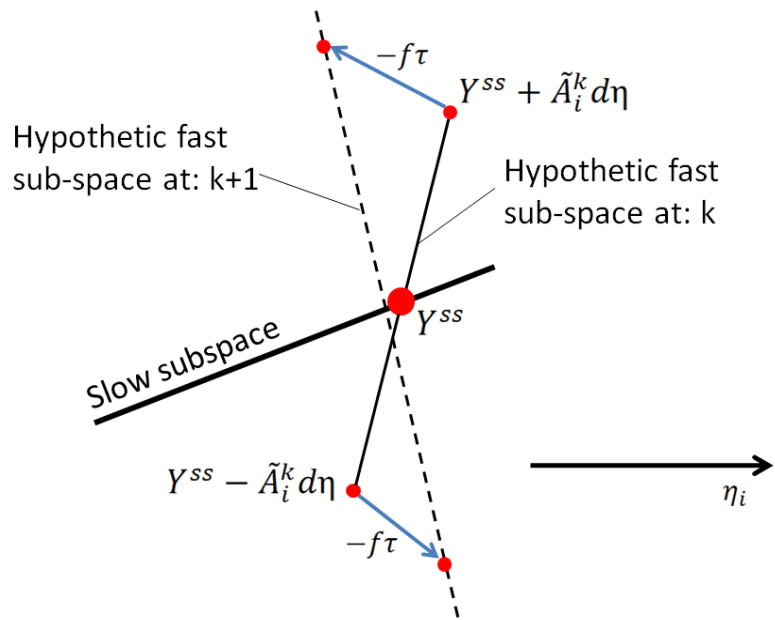}
	\caption{(Color online). {Rationale behind the} refinement process of the fast subspace: In a neighborhood of the SIM (slow subspace), the anti-parallel dynamics $-f$ reacts with a {\em torque} if (\ref{Y.fast}) does not span the fast subspace. As a result, the latter subspace is the {\em stable stationary} solution of Eqs. (\ref{A.fast.updated}) and (\ref{A.fast.dynamical}). {Small circles denote neighbors of a pivot (large circle).}}\label{schizzo.fast}
\end{figure} 

Let us assume that the fast subspace can be uniquely parameterized (at least locally) by the variables $\eta_i, \; i=1,...,n-q-r$. Let the $z \times n$ matrix $\tilde {B}$ and its $j$-th row $\tilde{B}_j$ 
\begin{equation}\label{Bmatrix.fast}
\tilde{B} = \left[ {\begin{array}{*{20}c}   {b_{11} } &  \cdots  & {b_{1n} }  \\    \vdots  &  \ddots  &  \vdots   \\   {b_{z1} } &  \cdots  & {b_{zn} }  \\\end{array}} \right],\; \tilde{B}_j  = \left[ {b_{j1} ,...,b_{jn} } \right], \quad z=n-q-r,
\end{equation}
define a linear mapping:
\begin{equation}\label{B.eta.fast}
\tilde{B}Y = \eta ,\quad \eta  = \left[ {\begin{array}{*{20}c}
   {\eta_1 }  \\
    \vdots   \\
   {\eta_z }  \\
\end{array}} \right],
\end{equation}
where the dynamics $f$ of the system (\ref{ODEgen}) obeys a set of $r$ linear conservation laws.
In a neighborhood of the SIM point $Y^{ss}$, at a given iteration $k$, the fast sub-space can be represented by a linear function as follows:
\begin{equation}\label{Y.fast}
Y  = \tilde{A}^k   \eta  + \tilde{l}^k,
\end{equation} 
with
\begin{equation}\label{AAl.fast}
\tilde{A}^k  = \left[ {\begin{array}{*{20}c}   {\tilde a_{11}^k } &  \cdots  & {\tilde a_{1z} ^k}  \\    \vdots  &  \ddots  &  \vdots   \\   {\tilde a_{n1}^k } &  \cdots  & {\tilde a_{nz}^k }  \\\end{array}} \right], \; \tilde A_i^k  = \left[ {\begin{array}{*{20}c}   {\tilde a_{1i} ^k}  \\    \vdots   \\   {\tilde a_{ni} ^k}  \\\end{array}} \right],\;\; \tilde l^k  = \left[ {\begin{array}{*{20}c}   {\tilde l_1 ^k}  \\    \vdots   \\   {\tilde l_n ^k}  \\\end{array}} \right].
\end{equation}
Similarly to the procedure of section \ref{slow.equations}, here we aim at devising an iterative procedure
\begin{equation}\nonumber
\tilde A^k \rightarrow \tilde A^{k+1}, \; \tilde l^k \rightarrow \tilde l^{k+1},
\end{equation} 
so that the function (\ref{Y.fast}), in the limit $k \rightarrow \infty$, accurately describes the fast subspace. Let $d \eta$ denote {an arbitrary} small parameter. Following the pictorial representation of Fig. \ref{schizzo.fast}, for every variable $\eta_i$ we consider the relaxation of the two neighbors (in the affine space (\ref{Y.fast})) of $Y^{ss}$
\begin{equation}\nonumber
Y^{ss} + \tilde A_i^k d \eta, \; Y^{ss} - \tilde A_i^k d \eta,
\end{equation}
under the anti-parallel dynamics $-f$. After time $\tau$, these states move to the new locations:
\begin{equation}\nonumber
Y^{ss}+\tilde A_i^k d \eta - f \left( Y^{ss} + \tilde A_i^k d\eta \right) \tau, \; Y^{ss}-\tilde A_i^k d \eta - f \left( Y^{ss} - \tilde A_i^k d\eta \right) \tau,
\end{equation}
which, upon linearization of the vector field $f$, take the approximate form:
\begin{equation}\nonumber
Y^{ss}+\tilde A_i^k d \eta - f\left( Y^{ss} \right) \tau - J\left( Y^{ss} \right) \tilde A_i^k d\eta  \tau, \; Y^{ss}-\tilde A_i^k d \eta - f\left( Y^{ss} \right) \tau + J\left( Y^{ss} \right) \tilde A_i^k d\eta  \tau.
\end{equation}
%
Therefore, a set of $z$ vectors spanning the fast sub-space at the iteration $k+1$ reads:
\begin{equation}\nonumber
\tilde A_i^k - J^{ss} \tilde A_i^k \tau, \quad i=1,...,z,
\end{equation}
where, for the sake of notations, $J^{ss}=J\left( Y^{ss} \right)$. We can thus describe the updated fast sub-space as follows:
\begin{equation}\nonumber
Y-Y^{ss}= \sum_{i=1}^{z} \tilde{\alpha}_i \left(\tilde A_i^k - J^{ss} \tilde A_i^k \tau \right),
\end{equation}
or equivalently in matrix notations
\begin{equation}\label{Function.fast}
Y-Y^{ss}=\tilde{M}^k \tilde{\Lambda},
\end{equation}
with
\begin{equation}\label{M.fast}
\tilde M^k = \left[ {\tilde A_1^k - J^{ss} \tilde A_1^k \tau}  ,    ...   ,   {\tilde A_z^k - J^{ss} \tilde A_z^k \tau}   \right], \tilde{\Lambda}=\left[\tilde \alpha_1,...,\tilde \alpha_z \right]^T.
\end{equation}
By substituting the (\ref{Function.fast}) into (\ref{B.eta.fast}),
\begin{equation}\label{eta.eq.01}
\eta - \tilde{B} Y^{ss}=\tilde{\Phi}_k \tilde \Lambda,
\end{equation}
where the generic element $\tilde \Phi_k \left( i,j \right)$ of the $z \times z$ matrix $\tilde \Phi_k$ reads
%
\begin{equation}\nonumber
\tilde{\Phi}_k \left( i,j \right) = \delta_{ij} - \tilde{B}_i J^{ss} \tilde{A}_j^k \tau,
\end{equation}
where, owing to (\ref{B.eta.fast}) and (\ref{Y.fast}), $\tilde{B}_i \tilde{A}_j^k = \delta_{ij} \; \forall k.$
%
From (\ref{Function.fast}) and (\ref{eta.eq.01}), it follows that at the iteration $k+1$ the linear function describing the fast sub-space is:
\begin{equation}\nonumber
Y=\tilde M \tilde \Phi_k^{-1} \left(\eta - \tilde B Y^{ss} \right) + Y^{ss},
\end{equation}
so that
\begin{equation}\label{A.fast.updated}
\tilde A^{k+1} = \tilde M^k \tilde \Phi_k^{-1}.
\end{equation}
Similarly to the (\ref{Akp1}), the updating rule (\ref{A.fast.updated}) can be regarded as the explicit Euler scheme for integrating the dynamical system:
\begin{equation}\label{A.fast.dynamical}
\frac{d \tilde A}{dt}=\left[\tilde M \tilde \Phi^{-1} - \tilde A \right] \tau^{-1},
\end{equation}
with
%
\begin{equation}\nonumber
\tilde M = \left[ {\tilde A_1 - J^{ss} \tilde A_1 \tau}  ,    ...   ,   {\tilde A_z - J^{ss} \tilde A_z \tau}   \right], \; \tilde{\Phi} \left( i,j \right) = \delta_{ij} - \tilde{B}_i J^{ss} \tilde{A}_j \tau .
\end{equation}
%
%
{\subsection{Stationary solutions of the linearized RRM equations}\label{steady.state.accuracy}
We notice that the matrix $A$ keeps evolving under the dynamics (\ref{Y.evol}) until the following steady condition holds: 
\begin{equation}\label{stationary.A01}
\frac{dA}{dt} = M \Phi^{-1} - A = 0,
\end{equation}
which can be recast in the more explicit form:
\begin{equation}\label{stationary.A}
(I + J \tau) A (I + B J A \tau)^{-1} = A.
\end{equation}
It is straightforward to prove that right eigenvectors of the Jacobian $J$ satisfy the stationary condition (\ref{stationary.A}). Let the columns of $A$ represent a set of $q$ eigenvectors of $J$ such that 
\begin{equation}\label{eigen.def}
J A = A L, 
\end{equation}
where $L$ is a $q \times q$ diagonal matrix whose non-zero components are the corresponding eigenvalues. Upon substitution of (\ref{eigen.def}) in (\ref{stationary.A}), we obtain the identity:
\begin{equation}\label{identity}
A(I + L \tau) (I + B A L \tau)^{-1} = A,
\end{equation} 
due to the condition $BA=I$. Similar considerations apply to the evolution of $\tilde A$ under the (\ref{A.fast.dynamical}). Hence, we can conclude that the eigenvectors of $J$ (evaluated at $Y^{ss}$) do provide stationary solution for both the equations $dA/dt=0$ and $d \tilde A /dt=0$. The identity (\ref{identity}) also suggests that, if (\ref{eigen.def}) holds, the projector operator in the first equation of (\ref{Y.evol}) takes the simple stationary form $P = M \Phi^{-1} B = A B$, such that the pivot evolution is ruled by:
\begin{equation}\label{Y.evol.ss}
\frac{dY}{dt}=(I-P)f=(I-AB)f.
\end{equation}
The above considerations suggest that the proposed method can deliver approximations of the SIM up to an accuracy of the order of ILDM \cite{MP92}.} 

{It is worth stressing that such a limit is not due to the RRM approach \cite{PHD_THESIS_EC,ChKa2011,CKPRE2010}, rather to the linear approximations of (\ref{mapping}) and (\ref{relaxed.points}) by (\ref{linearized.manifold}) and (\ref{relaxed.points.approx}), respectively. Hence, other governing equations leading to more accurate description of the SIM compared to (\ref{Y.evol}) may be also devised, abandoning the present linear expressions (\ref{linearized.manifold}) and (\ref{relaxed.points.approx}) in favor of higher order approximations (at the cost of a more demanding implementation).}
\subsection{Initialization}\label{inizio}
\begin{figure}
	\centering
		\includegraphics[width=1\textwidth]{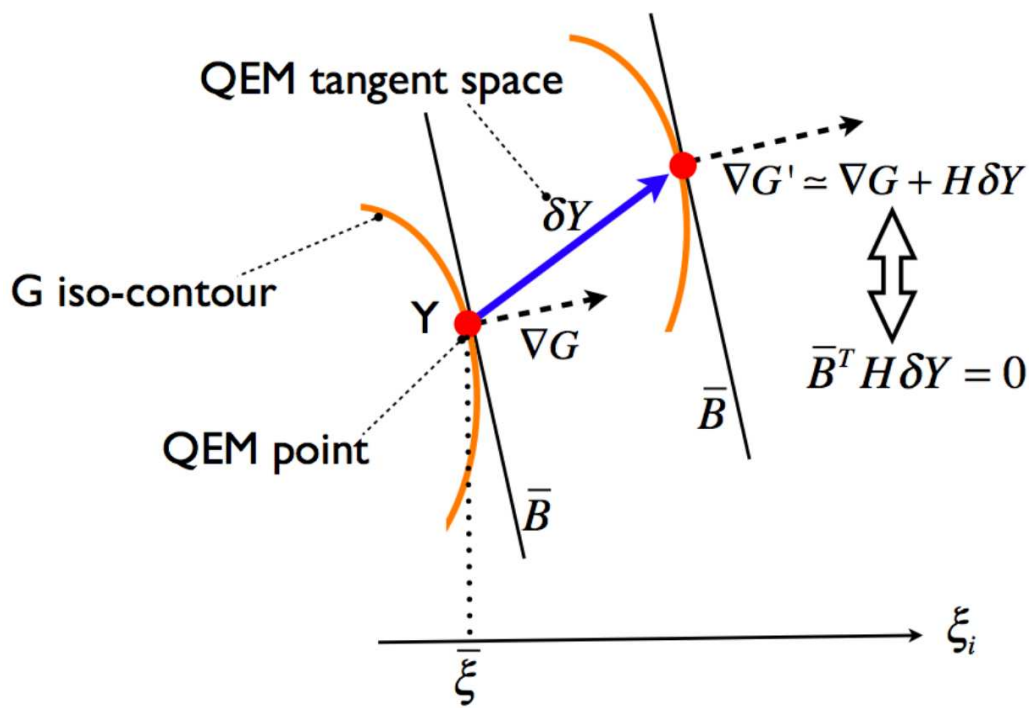} 
	\caption{(Color online). Pictorial representation of the notion of quasi-equilibrium manifold (QEM) (\ref{QEM.def}), $H$ and $\bar B$ being the second derivative matrix of the Lyapunov function $G$ and the null space of the full set of constraints in (\ref{QEM.def}), respectively.}\label{schizzo.QEM}
\end{figure}

The method described in section \ref{slow.equations} for constructing local approximations of SIM requires the initial choice of $Y^1$ and $A^1$.
Several approximations of the SIM can be adopted for this purpose as proposed in \cite{Lebiedz04,LEB-JCP}. In the following, we discuss in detail another possible initialization strategy for the case of dissipative systems. Closed chemically reactive mixture of gases are prototypical examples of large dissipative systems that can be addressed by model reduction techniques \cite{GoussisMaas}. In fact, due to the second law of thermodynamics, the dynamical system (\ref{ODEgen}), describing the temporal evolution of chemical species concentrations, is equipped with a thermodynamic Lyapunov function $G$ \cite{Lya} related to entropy and always decreasing in time. In this case, a rough approximation of the SIM is often provided by the {\em quasi-equilibrium manifold} (QEM) \cite{book,ChGoKa07,ChK2007}, also referred to as {\em constrained equilibrium manifold} \cite{KeG71,BisHam01,RenPope06}. A QEM is defined by means of the following constrained optimization problem:
\begin{equation}\label{QEM.def}
\left\{ \begin{array}{l}
 G\left( Y \right) \to \min,  \\ 
 BY = \left[ {\bar \xi _1 ,...,\bar \xi _q } \right]^T,  \\ 
 DY = \left[ {\chi _1 ,...,\chi _r } \right]^T.  \\ 
 \end{array} \right.
\end{equation}
where $q$ denotes the QEM dimension, while the $r \times n$ matrix $D$ imposes the conservation of the number of moles ($\chi_{i, i=1,...,r}$) of $r$ chemical elements involved in the reaction. 
In Fig. \ref{schizzo.QEM}, the geometry behind the notion of QEM is shown schematically. Let $H$ and $\bar B$ be the second derivative matrix of the Lyapunov function $G$ and the null space of the $(q+r) \times n$ matrix of contraints in (\ref{QEM.def}):
\begin{equation}\nonumber
\bar B = \ker \left[ \begin{array}{l}
 B_1  \\ 
  \vdots  \\ 
 B_q  \\ 
 D_1  \\ 
  \vdots  \\ 
 D_r  \\ 
 \end{array} \right],\quad D = \left[ \begin{array}{l}
 D_1  \\ 
  \vdots  \\ 
 D_r \\ 
 \end{array} \right] .
\end{equation}
Let the $ n \times (n-r) $ matrix $\bar D$ be defined as follows:
\begin{equation}\nonumber
 \bar D = \left[ {\bar D_1 ,...,\bar D_{n - r} } \right]
= \ker D.
\end{equation}
With the basis vectors $\{\bar D_1,...,\bar D_{n-r}\}$ spanning the null space of $D$ ($\ker D$), an arbitrary vector $\delta Y$ along the tangent space of a QEM can be written in terms of the vector $\delta=\left[ {\delta _1 ,...,\delta _{n - r} } \right]^T$ as:
\begin{equation}\label{D.delta}
\bar D \delta = \delta Y = \bar D\left[ \begin{array}{l}
 \delta _1  \\ 
  \vdots  \\ 
 \delta _{n - r}  \\ 
 \end{array} \right] .
\end{equation}
The geometry behind the optimization problem (\ref{QEM.def}) imposes the orthogonality condition $\bar B^T H \bar D \delta =0$ (see also Fig. \ref{schizzo.QEM}), and the tangent space to a QEM is spanned by:
\begin{equation}\nonumber
T_{QEM} = \bar D \ker (\bar B^T H \bar D). 
\end{equation}
Recalling the definition of the matrix $A$ in (\ref{linearized.manifold}), a possible initialization of $A$ (with $\mathcal{F}^1 : \quad Y  = A^1   \xi  + l^1$ describing locally a QEM) takes the form:
\begin{equation}\label{SLOPE_QEM}
A^1  = T_{QEM} \left( {B T_{QEM} } \right)^{ - 1} ,
\end{equation}
whereas $\mathcal{F}^1\left( \bar \xi \right)=Y^1$ can be found by solving the optimization problem (\ref{QEM.def}) using for example tools suggested in \cite{Reyn86} and \cite{ChK2007}. 
Finally, possible choices for the matrix $B$ are discussed in \cite{ChGoKa07,ChK2007} (spectral quasi equilibrium parameterization) and \cite{KeG71,BisHam01} (constrained equilibrium parameterization) while exact formulae for computing matrices $H$ and $J$ can be found in \cite{PHD_THESIS_EC,CF2010}.
%

Moreover, Eq. (\ref{A.fast.updated}) and the dynamical system (\ref{A.fast.dynamical}) require the initial condition $\tilde A^1=\tilde A(t_o=0)$. A possible option is the following: Since fast motions are necessarily transversal to the slow subspace (\ref{SIM.linear}), a reasonable choice for the $(n-q-r) \times n$ matrix $\tilde B$ reads:
\begin{equation}\label{fast.B.matrix}
\tilde B =\left( \ker \left[ \begin{array}{l}
 A^{ss}_1  \\ 
  \vdots  \\ 
 A^{ss}_q  \\ 
 D_1  \\ 
  \vdots  \\ 
 D_r  \\ 
 \end{array} \right] \right)^T.
\end{equation}
As a first guess ($k=1$), let the mapping (\ref{Y.fast}) describe the orthogonal subspace to the SIM (\ref{SIM.linear}). More specifically, let the former space be spanned by the columns of the $n \times (n-q-r)$ matrix
\begin{equation}\label{T.ortho}
T^\bot= \bar D  \ker \left[ (A^{ss})^T \bar D\right],
\end{equation}    
similarly to (\ref{SLOPE_QEM}), initial condition for (\ref{A.fast.updated}) and (\ref{A.fast.dynamical}) takes the explicit form
\begin{equation}\label{IC.fast}
\tilde A^1 = T^\bot \left(\tilde B T^\bot\right)^{-1}.
\end{equation}
%
We notice that owing to the relations $B A^1 = I$ and $\tilde{B} \tilde{A}^1 = I$, as an alternative to (\ref{SLOPE_QEM}) and (\ref{IC.fast}), $A$ and $\tilde{A}$ can be initialized by computing the Moore-Penrose pseudoinverse matrices of $B$ and $\tilde{B}$, respectively.
%
\subsection{Stability and adaptive construction of slow invariant manifolds}
Stability of the governing equations of the linearized RRM (\ref{Y.evol}) can be exploited for adaptive construction of SIM. In the first place, Eqs. (\ref{Y.evol}) can be solved with $q=1$: If convergence is experienced, we assume that a 1D reduced model of the system (\ref{ODEgen}) can be constructed in a vicinity of the macrostate $\bar \xi$. In other words, a {\em minimal description} (\ref{ODEred}) of the detailed system (\ref{ODEgen}) can be accomplished by means of one degree of freedom. On the contrary, with no convergence, the manifold dimension is updated to $q=2$ and the procedure repeated. Upon convergence with some $q = \bar q$, we may infer that a {\em minimal description} of the detailed dynamics requires $\bar q$ degrees of freedom. In this sense, the suggested method enables an adaptive construction of SIM (i.e. varying dimension in the phase-space without any {\em a priori} assumptions on the value of $q$). The above idea relies upon the assumption that RRM is stable provided the existence of SIM of a certain dimension $q$ \cite{ChKa2011,CKPRE2010}. More details on the stability of the RRM can be found in \cite{PHD_THESIS_EC}, where a comparative study between a method for the direct solution of the film equation (\ref{film}) and RRM is performed.
\section{Benchmark}\label{BEN-EX}
\begin{figure}
	\centering
		\includegraphics[trim = 20mm 3mm 15mm 0mm, clip, width=1\textwidth]{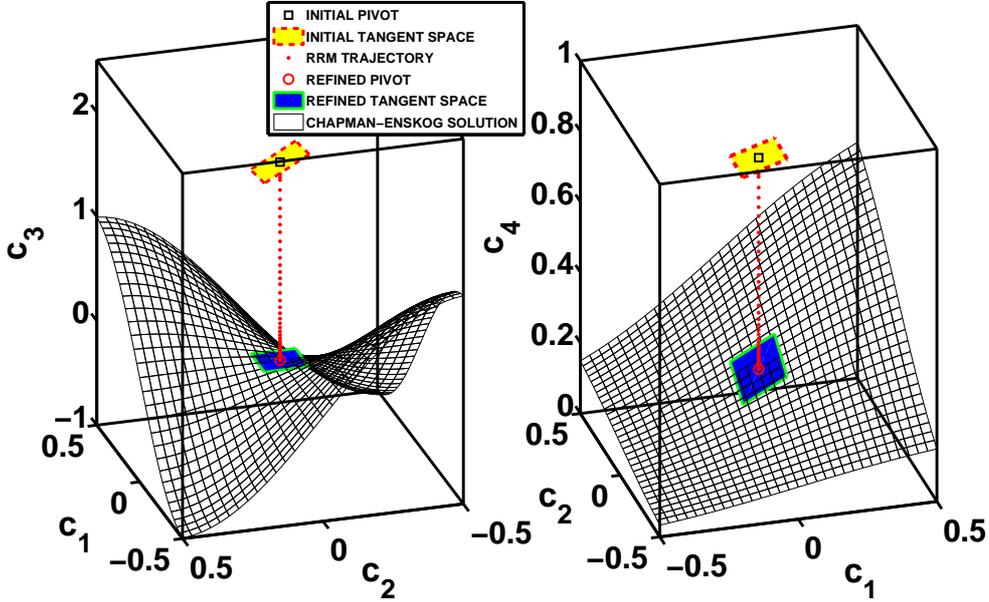}
	\caption{(Color online). Slow invariant manifold with respect to the dynamical system (\ref{ODEben}) with (\ref{choice}), $\omega=3$ and $\epsilon=0.025$. Starting from the initial conditions (\ref{initial.conditions}), the governing equations of the linearized RRM (\ref{Y.evol}) with $n=4$, $q=2$ and $\tau=3 \times 10^{-10}$ are solved by means of the stiff numerical scheme {\em ode15s} readily available in Matlab$\textsuperscript{\textregistered}$ \cite{ode15s}. Here, the steady state is reached after {an integration} time $T_f=1$. 
At steady state, the solution trajectory (dots) finally {\em lands} on the SIM.}\label{CASO-BEN01}
\end{figure} 
For the sake of simplicity, we consider here a four-dimensional model where the dynamics of two fast variables ($c_3$ and $c_4$) is {\em slaved} to the motion of the slow variables $c_1$ and $c_2$ \cite{ben}. Let the functions $f_1$, $f_2$, $\theta_1$ and $\theta_2$ depend on $c_1$ and $c_2$ only. In the following, we focus on the ODE system:
\begin{equation}\label{ODEben}
\frac{{dc}}{{dt}} = \left[ {\begin{array}{*{20}c}
   {f_1 \left( c_1,c_2 \right)}  \\
   {f_2 \left( c_1,c_2 \right)}  \\
   {-\frac{1}{\epsilon} \left[c_3 - \theta_1 \left( c_1,c_2 \right) \right] + f_1 \partial_{c_1} \theta_1 \left( c_1,c_2 \right) + f_2 \partial_{c_2} \theta_1 \left( c_1,c_2 \right) }  \\
   {-\frac{1}{\epsilon} \left[c_4 - \theta_2 \left( c_1,c_2 \right) \right] + f_1 \partial_{c_1} \theta_2 \left( c_1,c_2 \right) + f_2 \partial_{c_2} \theta_2 \left( c_1,c_2 \right) }  \\
\end{array}} \right],
\end{equation}
\begin{figure}
	\centering
		\includegraphics[trim = 5mm 1mm 15mm 0mm, clip, width=1\textwidth]{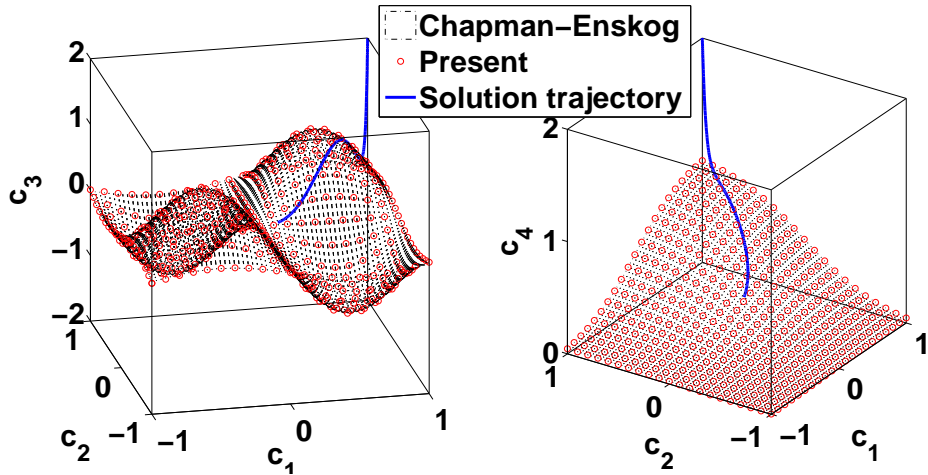}
	\caption{(Color online). An array of initial states has been refined by means of the linearized RRM. Stationary states of the (\ref{Y.evol}) are reported (circles). For a comparison, the exact Chapman-Enskog solution and a detailed solution trajectory of the system (\ref{ODEben}) are also shown. Here we use $\omega=3$, $\tau=3 \times 10^{-10}$, while integration of (\ref{Y.evol}) is performed for {an integration} time $T_f=1$ at any point. The computational time required to refine the entire array, composed by $(21 \times 21)$ points, was $2.5$ minutes using a Matlab code on a single processor with 1.73 GHz.}\label{glob}
\end{figure} 
where $\epsilon$ and $\partial_i$ denote a fixed small quantity and partial derivative with respect to variable $i$, respectively. Assuming that the dynamics of $c_3$ and $c_4$ is slaved to the slow variables,
\begin{equation}\label{slaving}
c_3=c_3\left( c_1,c_2 \right), \; c_4=c_4\left( c_1,c_2 \right).
\end{equation}
and, according to the chain rule, time derivatives of (\ref{slaving}) take the explicit form:
\begin{equation}\label{chain_rule}
{\begin{array}{*{20}c}
\frac{dc_3}{dt}= \partial_{c_1} c_3 \frac{dc_1}{dt} + \partial_{c_2} c_3 \frac{dc_2}{dt} = \partial_{c_1} c_3 f_1 + \partial_{c_2} c_3 f_2,\\
\frac{dc_4}{dt}= \partial_{c_1} c_4 \frac{dc_1}{dt} + \partial_{c_2} c_4 \frac{dc_2}{dt} = \partial_{c_1} c_4 f_1 + \partial_{c_2} c_4 f_2.\\
\end{array}}
\end{equation}
\begin{figure}
	\centering
		\includegraphics[trim = 10mm 2mm 30mm 1mm, clip, width=1\textwidth]{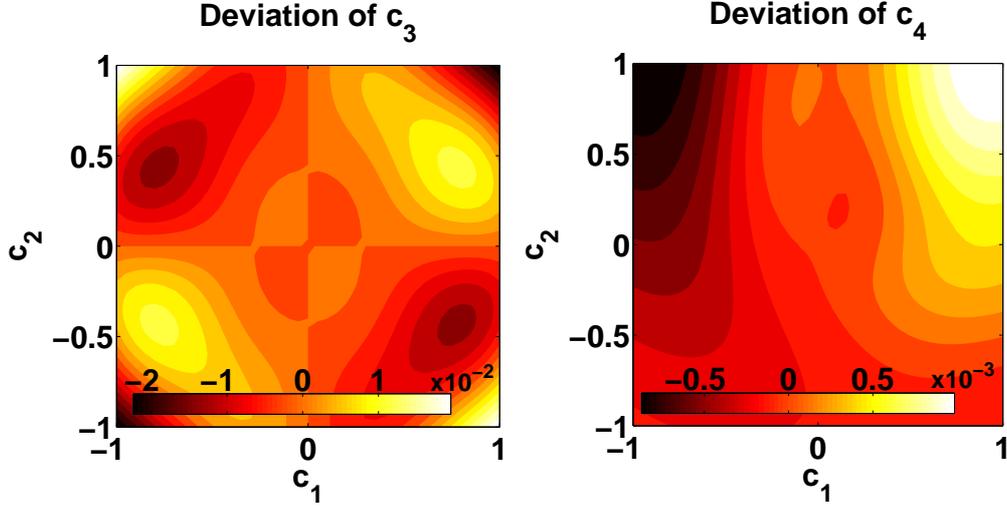}
	\caption{(Color online). Difference between the SIM constructed by the linearized RRM method as reported in Fig. \ref{glob} and the Chapman-Enskog solution ($c_3=\theta_1$, $c_4=\theta_2$).}\label{Deviation}
\end{figure}
Upon substituting equations (\ref{chain_rule}) in (\ref{ODEben}), one obtains the following {\em invariance conditions} with respect to (\ref{ODEben}) (see also section \ref{background} and Eq. (\ref{invEq})):
\begin{equation}\label{invariance.conditions}
{\begin{array}{*{20}c}
{\partial_{c_1} c_3 f_1 + \partial_{c_2} c_3 f_2 = -\frac{1}{\epsilon} \left[c_3 - \theta_1 \left( c_1,c_2 \right) \right] + f_1 \partial_{c_1} \theta_1 \left( c_1,c_2 \right) + f_2 \partial_{c_2} \theta_1 \left( c_1,c_2 \right),} \\
{\partial_{c_1} c_4 f_1 + \partial_{c_2} c_4 f_2 = -\frac{1}{\epsilon} \left[c_4 - \theta_2 \left( c_1,c_2 \right) \right] + f_1 \partial_{c_1} \theta_2 \left( c_1,c_2 \right) + f_2 \partial_{c_2} \theta_2 \left( c_1,c_2 \right).} \\
\end{array}}
\end{equation}
A common approach to obtain solutions to the above invariance conditions is the {\em Chapman-Enskog method} \cite{CEexpansion,Struchtrupbook}, which is based on the assumption that $\epsilon$ is small compared to all other quantities, and it is implemented by series expansions of the (\ref{slaving}) in powers of $\epsilon$:
\begin{equation}\label{expansion}
{\begin{array}{*{20}c}
c_3\left( c_1,c_2 \right) = c_3^{(0)} + \epsilon c_3^{(1)} + \epsilon^2 c_3^{(2)} + ...\\
c_4\left( c_1,c_2 \right) = c_4^{(0)} + \epsilon c_4^{(1)} + \epsilon^2 c_4^{(2)} + ...\\
\end{array}}
\end{equation}
Hence, the first equation in (\ref{invariance.conditions}) reads:
\begin{equation}\nonumber
\begin{split}
f_1 \partial_{c_1} \left[ c_3^{(0)} + \epsilon c_3^{(1)} + \epsilon^2 c_3^{(2)} + ... \right] + f_2 \partial_{c_2} \left[ c_3^{(0)} + \epsilon c_3^{(1)} + \epsilon^2 c_3^{(2)} + ... \right] = \\ 
-\frac{1}{\epsilon} \left[c_3^{(0)} + \epsilon c_3^{(1)} + \epsilon^2 c_3^{(2)} + ... - \theta_1  \right] + f_1 \partial_{c_1} \theta_1  + f_2 \partial_{c_2} \theta_1 .
\end{split}
\end{equation}
After collecting terms with the same power of $\epsilon$, we obtain:
\begin{equation}\label{CHAP_sol}
c_3^{(0)} = \theta_1 \left( c_1,c_2 \right), \; c_3^{(i)} = 0, \; \forall i>0,
\end{equation}
namely $c_3 = \theta_1 \left( c_1,c_2 \right)$, and similarly $c_4 = \theta_2 \left( c_1,c_2 \right)$.

\begin{figure}
	\centering
		\includegraphics[trim = 5mm 3mm 15mm 0mm, clip, width=1\textwidth]{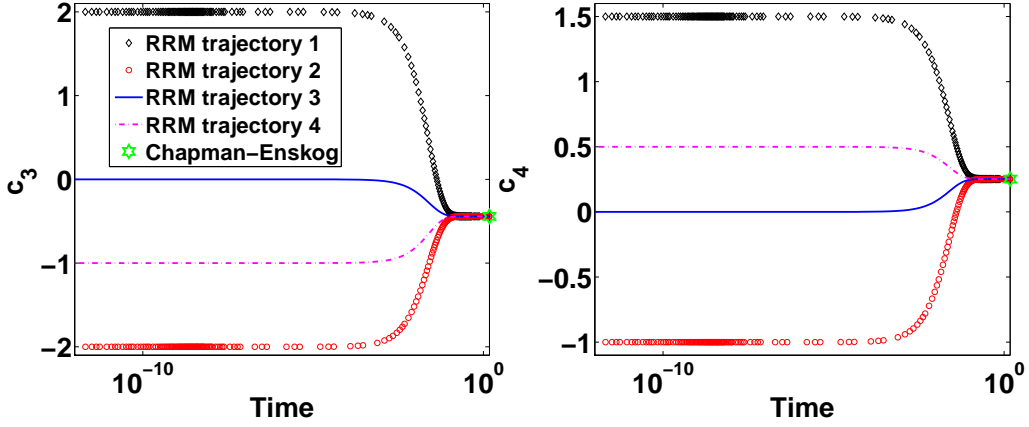}
	\caption{(Color online). On the basis of the system (\ref{ODEben}) with (\ref{choice}), $\omega=3$ and $\epsilon=0.025$, we solve (\ref{Y.evol}) by means of the numerical scheme {\em ode15s} \cite{ode15s} starting from several initial conditions $\bar Y^1$ and $A^1$, with fixed parameters $\xi_1=c_1=0.3$ and $\xi_2=c_2=-0.2$. Time evolution of $c_3$ and $c_4$, as dictated by (\ref{Y.evol}), is reported. {For an integration} time $T_f > 1$, the steady state is reached such that $c_3^{ss}=-0.4422$, $c_4^{ss}=0.2520$ for any initial conditions. At steady state, small deviations from the Chapman-Enskog solution (star) are observed (smaller than 0.01\%).}\label{CASO-BEN02}
\end{figure}

For illustration purposes, we choose: 
\begin{equation}\label{choice}
{\begin{array}{*{20}l}
\theta_1 \left( c_1 , c_2 \right)={\rm sin}\left(\omega c_1\right) {\rm sin} \left(\omega c_2 \right), \; 
\theta_2 \left( c_1 , c_2 \right) = \left[ \left( 1 + e^{-\omega c_1} \right) \left( 1 + e^{-\omega c_2} \right) \right]^{-1}, \\
f_1=-c_1, \; f_2=-2c_2.
\end{array}}
\end{equation}
In Fig. \ref{CASO-BEN01}, the Chapman-Enskog solution to (\ref{invariance.conditions}) is plotted to illustrate the relaxation of system (\ref{Y.evol}), starting from {the following} initial pivot and tangent space:
\begin{equation}\label{initial.conditions}\bar Y^1  = \left[ {0,0,1.9,0.85} \right]^T ,\;\;A^1  = \left[ {\begin{array}{*{20}c}   1 & 0  \\   0 & 1  \\   { - 0.276} & { - 1.405}  \\   {0.225} & {0.0282}  \\\end{array}} \right].
\end{equation}
Finally, the SIM parameterization is chosen as follows: $\xi_1=c_1$, $\xi_2=c_2$. We stress that, in the computations, the steady state ($\bar Y^{ss}$, $A^{ss}$) does not depend on the initial choice of $\bar Y^1$ and $A^1$ (see Fig. \ref{CASO-BEN02}). The latter observation is consistent with the idea behind the RRM \cite{CKPRE2010}, which can be elucidated by saying that states on the SIM represent {\em stable steady states} of the dynamical system (\ref{Y.evol}).   

The reduced system (\ref{ODEred}) for the above example (\ref{ODEben}) {rules the evolution of the slow variables}:
\begin{equation}\label{reducedODE}
{\begin{array}{*{20}l}
\frac{dc_1}{dt}=f_1=-c_1\\
\frac{dc_2}{dt}=f_2=-2c_2
\end{array}}
\end{equation}
whereas fast variables can be reconstructed by means of the mappings $c_3=c_3(c_1,c_2)$, $c_4=c_4(c_1,c_2)$ (Fig. \ref{glob}). 

\begin{figure}
	\centering
		\includegraphics[trim = 10mm 0mm 15mm 0mm, clip, width=1\textwidth]{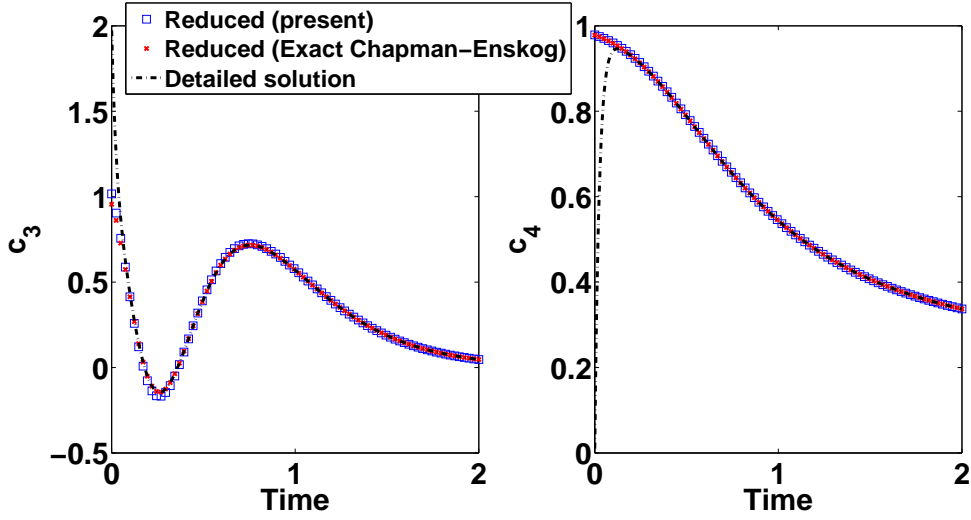}
	\caption{(Color online). Comparison between the detailed solution of the system (\ref{ODEben}) from the state $c_1=1.5$, $c_2=1.5$, $c_3=2$, $c_4=0$ (out of SIM) and the reduced solution of the system (\ref{reducedODE}) from the state $c_1=1.5$, $c_2=1.5$. The numerical stiff solver {\em ode15s} \cite{ode15s} is used with $\omega=3$, $\epsilon=0.025$ and $\tau=3 \times 10^{-10}$.}\label{relax.comp}
\end{figure}

In this respect, in Fig. \ref{relax.comp}, a solution trajectory of (\ref{ODEben}) is compared to the trajectory of (\ref{reducedODE}), where the reconstruction of $c_3$ and $c_4$ is performed using both the exact Chapman-Enskog solution ($c_3=\theta_1$, $c_4=\theta_2$) and a linear look up table based on the nodes refined by the linearized RRM {(see also Figs. \ref{glob} and \ref{Deviation}).} 
\begin{table}
\begin{center}
    \begin{tabular}{ | c | c | c | c | c |}
    \hline
    $\tau$ & $c_1^{ss}$ & $c_2^{ss}$ & $c_3^{ss}$ & $c_4^{ss}$ \\ \hline
    $1 \times 10^{-13}$ & $-0.6$ & $-0.85$ & ${0.551271605706663}$ & ${0.010023694645548}$ \\
    $1 \times 10^{-12}$ & $-0.6$ & $-0.85$ & ${0.551271602038736}$ & ${0.010023695783466}$ \\
    $1 \times 10^{-11}$ & $-0.6$ & $-0.85$ & ${0.551271598194113}$ & ${0.010023697018395}$ \\
    $1 \times 10^{-10}$ & $-0.6$ & $-0.85$ & ${0.551271596762428}$ & ${0.010023697470180}$ \\
    $1 \times 10^{-9}$  & $-0.6$ & $-0.85$ & ${0.551271605620228}$ & ${0.010023694672833}$ \\
    $1 \times 10^{-8}$  & $-$ & $-$ & $-$ & $-$ \\
    \hline
    \end{tabular}
    \caption{The steady state ($c_1^{ss}$, $c_2^{ss}$, $c_3^{ss}$, $c_4^{ss}$) of (\ref{Y.evol}) is computed for several choices of the parameter $\tau$ starting from the state $c_1=-0.6$, $c_2=-0.85$, $c_3=-1$, $c_4=0.5$, with $\omega=3$ and $\epsilon=0.025$. No convergence is observed for $\tau \ge 1 \times 10^{-8}$.}\label{Tabletau}
\end{center}
\end{table}

In Table \ref{Tabletau}, we report a sensitivity analysis with respect to the parameter $\tau$. 
We notice that an estimate of the time scales of a dynamical system can be obtained by a spectral decomposition of the Jacobi matrix. For the case in Fig. \ref{glob}, at equilibrium ($c_1=c_2=c_3=0$, $c_4=0.25$):
%
\begin{equation}\nonumber
\begin{split}
\tau_1=\tau_2=\frac{1}{\left|\lambda_1\right|}=\frac{1}{\left| \lambda_2 \right|}=3.72 \times 10^{-10}, \; \tau_3=\frac{1}{\left| \lambda_3 \right|}=7.45 \times 10^{-9}, \\  \tau_4=\frac{1}{\left| \lambda_4 \right|}=1.49 \times 10^{-8},
\end{split}
\end{equation}
%
where $\lambda_i$ denotes the $i$-th eigenvalue. Hence, in the above computations we use $\tau=3 \times 10^{-10}$. However, the latter parameters was varied within a wide range of values and no significant effect was noticed on both the stability of (\ref{Y.evol}) and the value of its steady state. 

In addition, we test the equations (\ref{A.fast.dynamical}) for computing the mapping (\ref{Y.fast}) describing the local fast subspace. To this end, we make use of (\ref{fast.B.matrix}) and (\ref{IC.fast}) (with $D=\bar D=1$, $r=0$).

At steady state of (\ref{A.fast.dynamical}), we observe (at any node of the grid in Fig. \ref{glob}) that the columns of the matrix $\tilde A$ span the subspace defined by the vectors:
\begin{equation}\nonumber\left[ {0,0,1,0} \right],\quad \left[ {0,0,0,1} \right],\end{equation}
in accordance with the assumption that $c_3$ and $c_4$ are the fast variables.
%
\begin{figure}
	\centering
		\includegraphics[trim = 16mm 3mm 15mm 0mm, clip, width=1\textwidth]{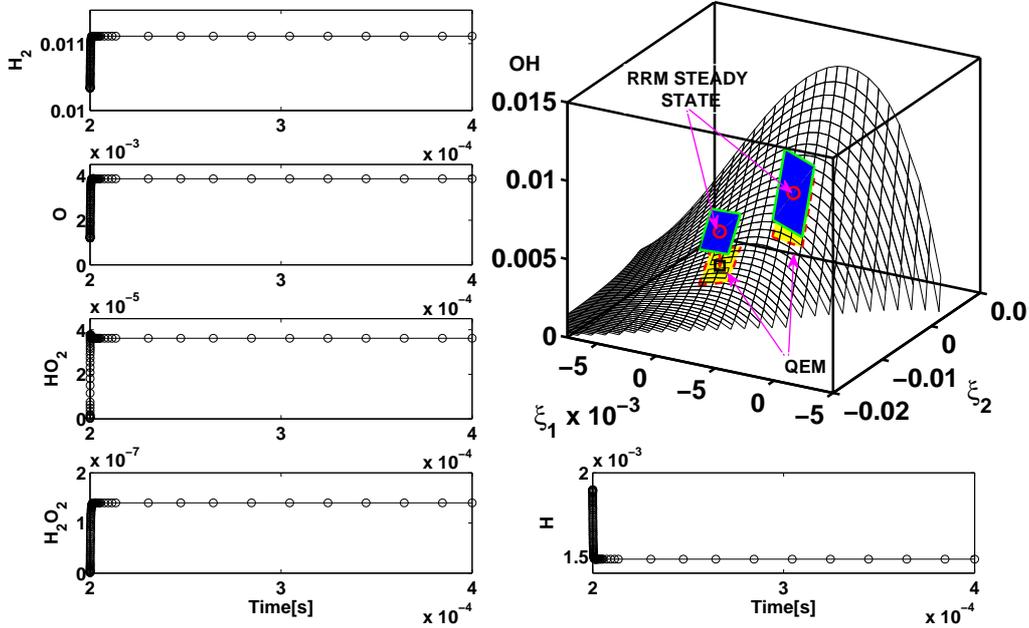}
	\caption{(Color online). Model for the hydrogen-air combustion \cite{Limech} in a closed system under fixed enthalpy ($500 kJ/kg$) and pressure ($1 bar$). Initialization is accomplished using (\ref{QEM.def}) and (\ref{SLOPE_QEM}), where the matrix $B$ is chosen according to the {\em spectral quasi equilibrium} parameterization \cite{ChGoKa07}. Here, an arbitrary state on a two dimensional QEM is driven on a two dimensional SIM by solving the governing equations of the linearized RRM (\ref{Y.evol}), with $n=9$, $q=2$, $r=3$, $\tau=1 \times 10^{-9}$ using the stiff numerical scheme {\em ode15s} readily available in Matlab$\textsuperscript{\textregistered}$ \cite{ode15s}.}\label{H2}
\end{figure} 

Finally, the governing equations of the linearized RRM (\ref{Y.evol}) were applied to a more complicated case of the detailed reaction mechanism for combustion of hydrogen in air \cite{Limech}. Here, Eq. (\ref{Y.evol}) were tested for computing states of the SIM with dimensions up to $q=5$. In Fig. \ref{H2}, we report an example with $q=2$.

Moreover, we observed that any steady state of (\ref{A.fast.dynamical}) corresponds to a matrix $\tilde A$ whose columns are linear combinations of the fast eigenvectors of $J^{ss}$ (i.e., eigenvectors corresponding to the $n-q-r$ largest eigenvalues in absolute value).
\section{Conclusions and outlook}\label{conclusions}
Based on the rationale behind the film equation (\ref{film}) and the Relaxation Redistribution Method (RRM), a set of ordinary differential equations (ODEs) is obtained with the aim of mimicking the only fast relaxation of a multiscale dynamical system towards a slow invariant manifold (SIM). This approach is characterized by a straightforward implementation consisting in solving the ODEs by state of the art stiff numerical schemes, and it proves useful for constructing accurate approximations of SIM in any dimensions. 

It is worth stressing that, like RRM, convergence of equations (\ref{Y.evol}) towards a steady state might be used for a fully adaptive construction of heterogeneous (i.e. varying dimension in different regions of the phase-space) slow invariant manifolds.

This work sheds light on the connection between the RRM method and the solution of both the invariance and film equations as postulated in \cite{PHD_THESIS_EC,ChKa2011,CKPRE2010} (see discussion at the end of section \ref{slow.equations}).

In addition, the novel algorithm (\ref{A.fast.dynamical}) for approximating the fast subspace is suggested, and a possible initialization procedure for both (\ref{Y.evol}) and (\ref{A.fast.dynamical}) is proposed. 

The methods are tested in the case of detailed combustion of hydrogen and air, as well as in a benchmark problem of a model with exact Chapman-Enskong solution of the invariance equation.

We stress that, although the presented methodology has been tested in the case of dissipative systems with a unique steady state (see Section \ref{BEN-EX}), in this paper we show that the governing equations (\ref{Y.evol}) and (\ref{A.fast.dynamical}) of the linearized RRM are based on the general notions of film equation (\ref{film}) and SIM (local) parameterization. Therefore, investigations on the performance of the presented method in more general systems with multiple steady states and chaotic behavior (see, e.g., \cite{Georgiou1999,Haller2010}) are planned for future publications. However, in the latter case, new initialization procedures are needed since the method discussed in Section \ref{inizio} is suitable for dissipative systems equipped with thermodynamic Lyapunov function.

{Finally, it is worth noticing that the proposed approach represents only one possible implementation of the RRM method (\cite{PHD_THESIS_EC,ChKa2011,CKPRE2010}). More accurate descriptions of the SIM (to be addressed in future publications) can be obtained as well, abandoning the present linear expressions (\ref{linearized.manifold}) and (\ref{relaxed.points.approx}) in favor of higher order approximations.}

\section{Acknowledgments}
I. Kevrekidis and B. Sonday are gratefully acknowledged for suggesting the model (\ref{ODEben}) with the functions (\ref{choice}). The Author wishes to thank S. Pope and I. Karlin for inspiring discussions. Acknowledgments go to C. Frouzakis and all Referees for their valuable help in improving the quality of the manuscript.

\bibliographystyle{plain}

\end{document}